\documentclass[journal]{IEEEtran}

\ifCLASSINFOpdf
\else
   \usepackage[dvips]{graphicx}
\fi
\usepackage{url}

\usepackage{graphicx}
\usepackage{multirow}
\usepackage{amsmath}
\usepackage{cite}
\usepackage{enumitem}
\usepackage{upgreek}
\usepackage{bm}
\usepackage{caption}
\usepackage{amssymb}
\usepackage{pifont}
\usepackage{comment}
\usepackage[font=scriptsize]{caption}

\begin{document}

\title{Knowledge-Decoupled Functionally Invariant Path with Synthetic Personal Data for Personalized ASR}

\author{Yue Gu, Zhihao Du, Ying Shi, Jiqing Han, \textit{Member, IEEE}, Yongjun He
\thanks{We sincerely thank Prof. Dong Wang (Tsinghua University) for valuable comments. This work was supported by NSFC under Grant 62376071. (Corresponding author: Zhihao Du, duzhihao.china@gmail.com; Jiqing Han, jqhan@hit.edu.cn.)}
\thanks{Yue Gu, Ying Shi, Jiqing Han, and Yongjun He are with the Research Center of Auditory Intelligence, School of Computer Science and Technology, Faculty of Computing, Harbin Institute of Technology, Harbin, China (e-mail: 427gy@sina.com; shiyingcs@hit.edu.cn; heyongjun@hit.edu.cn).}
\thanks{Zhihao Du is with the Speech Lab of Alibaba Group, Beijing, China.}}

\markboth{Journal of \LaTeX\ Class Files, Vol. 14, No. 8, August 2015}
{Shell \MakeLowercase{\textit{et al.}}: Bare Demo of IEEEtran.cls for IEEE Journals}
\maketitle

\begin{abstract}
Fine-tuning generic ASR models with large-scale synthetic personal data can enhance the personalization of ASR models, but it introduces challenges in adapting to synthetic personal data without forgetting real knowledge, and in adapting to personal data without forgetting generic knowledge. Considering that the functionally invariant path (FIP) framework enables model adaptation while preserving prior knowledge, in this letter, we introduce FIP into synthetic-data-augmented personalized ASR models. However, the model still struggles to balance the learning of synthetic, personalized, and generic knowledge when applying FIP to train the model on all three types of data simultaneously. To decouple this learning process and further address the above two challenges, we integrate a gated parameter-isolation strategy into FIP and propose a knowledge-decoupled functionally invariant path (KDFIP) framework, which stores generic and personalized knowledge in separate modules and applies FIP to them sequentially. Specifically, KDFIP adapts the personalized module to synthetic and real personal data and the generic module to generic data. Both modules are updated along personalization-invariant paths, and their outputs are dynamically fused through a gating mechanism. With augmented synthetic data, KDFIP achieves a 29.38\% relative character error rate reduction on target speakers and maintains comparable generalization performance to the unadapted ASR baseline.
\end{abstract}


\IEEEpeerreviewmaketitle

\section{Introduction}

\IEEEPARstart{P}{ersonalized} ASR models for individual speakers are essential in practice \cite{DBLP:journals/spl/LeeMKSC25}. To enhance acoustic-level personalization, speaker adaptation techniques adapt generic ASR models to specific speakers by eliminating the mismatch in voice characteristics between training and testing.

Depending on whether speaker-specific adaptation data are incorporated, speaker adaptation methods are generally classified into embedding-based and model-based ones. The former typically incorporates speaker embeddings \cite{DBLP:journals/taslp/DehakKDDO11,DBLP:conf/icassp/SnyderGSPK18,xue2014fast, DBLP:conf/interspeech/DelcroixWOKN18} or token embeddings \cite{li24ka_interspeech}, into the training of ASR models to obtain speaker-dependent ASR models \cite{saon2013speaker, DBLP:conf/icassp/SeniorL14, DBLP:conf/interspeech/ZeineldeenXLSN22, sari2020unsupervised, fan2019speaker, zhao2020speech, wan2020speaker}. Model-based methods fine-tune a generic ASR model \cite{DBLP:conf/interspeech/0028Y0021,gu23_interspeech} or speaker-specific parameters \cite{swietojanski2016learning, wang2017unsupervised, xie2021bayesian, deng23b_interspeech,gu24b_interspeech} with speaker-specific data, enabling a better capture of the target speaker's voice characteristics. Thus, these types of methods generally outperform embedding-based approaches \cite{bell2020adaptation}. Recently, fine-tuning limited speaker-specific parameters instead of the entire model, such as parameter-efficient fine-tuning (PEFT) \cite{houlsby2019parameter, DBLP:conf/iclr/HuSWALWWC22, hu2023llm, li2023evaluating}, holds a dominant position due to efficient adaptation. These methods follow a parameter-isolation \cite{DBLP:conf/acl/WangLJWWJCHWSZ23} strategy that keeps non-target and target speaker parameters separate. 

Although model-based speaker adaptation approaches have achieved certain success, the scarcity of speaker-specific data hinders further performance improvements, partly due to privacy concerns.
Empirical evidence indicates that generating additional personal speech data with broader textual coverage is an effective strategy for data augmentation and improves personalized ASR \cite{huang2020using, yang2023text, DBLP:conf/icassp/KimLC24}. Recently, large-scale zero-shot text-to-speech (TTS) models \cite{du2024cosyvoice, du2024cosyvoice2, du2025cosyvoice, anastassiou2024seed, le2023voicebox} have demonstrated human-level naturalness, expressiveness, and a diverse range of speaker profiles. These advances highlight their strong potential for data augmentation in low-resource ASR tasks \cite{yang2025enhancing}, i.e., augmenting the real data of the target speaker with synthetic data to enhance personalized ASR models with respect to the speaker's voice characteristics. However, hallucinations from large-scale TTS models \cite{liu2025mitigating} introduce phonetic or prosodic errors, potentially disturbing those model-based methods that rely on speaker-specific data. Moreover, when enhancing personalization with large-scale synthetic data, it is crucial for practical purposes to maintain generalization on non-target speakers, as considered in recent model-based adaptation methods \cite{gu23_interspeech,vander2023using,gu24b_interspeech}.

The key to addressing the above challenges lies in enabling the model to learn synthetic personal data without forgetting real knowledge and to learn personal data without forgetting generic knowledge. The two learning tasks can be modeled as new data adaptations under functional invariance. Given that the functionally invariant path (FIP) framework \cite{raghavan2024engineering} constructs a model updating path in the weight space that adapts to new data without impairing existing functionality, such as general recognition ability, we introduce FIP into personalized ASR tasks augmented with synthetic personal data.

When FIP is directly applied to personalized ASR, the model is trained on synthetic, real personal, and generic data, where large-scale synthetic data may compete with generic data, making it hard to balance different knowledge sources. To address this, we integrate the parameter-isolation strategy into FIP, storing generic and personalized knowledge in separate modules as in PEFT and applying FIP to them sequentially. To balance general representations from the generic module with speaker-specific information from the personalized module, we use the gating mechanism of our personality-memory gated adaptation (PGA) \cite{gu24b_interspeech} to dynamically fuse their outputs. Accordingly, we propose knowledge-decoupled FIP (KDFIP) for synthetic-data-augmented personalized ASR models. After storing generic and personalized knowledge separately, KDFIP updates each module along the gated personalization-invariant paths, fully leveraging synthetic personal data while preserving generalization.

\section{KDFIP with Synthetic Personal Data}
\noindent The schematic of the proposed KDFIP is shown in Fig.~\ref{fig:path}. Before detailing KDFIP, we first introduce synthetic data generation and the FIP framework to facilitate an intuitive understanding of the proposed method.
\vspace{-0.2cm}
\subsection{Data Augmentation with Zero-shot TTS Model}
We augment the personal data with the zero-shot large-scale TTS model, CosyVoice 2.0.
Our data augmentation process involves the following steps:
\begin{enumerate}[leftmargin=*]
\item Randomly select an utterance (and its paired text segment $\mathbf{y}_\mathrm{per}$) in a given personal corpus as the reference speech and extract speech tokens $\boldsymbol{\mu}_\mathrm{per}$, speaker embedding $\boldsymbol{v}$, Mel-filter bank feature $\boldsymbol{x}_\mathrm{per}$. Then, randomly select a text segment $\mathbf{y}_\mathrm{syn}$ from our internal multi-domain text database.
\item Construct the input sequence for text-to-token language model of CosyVoice 2.0 as ``\{$\mathbf{y}_\mathrm{per}$, $\mathbf{y}_\mathrm{syn}$, $\boldsymbol{\mu}_\mathrm{per}$\}'' and generate the speech tokens $\boldsymbol{\mu}_\mathrm{syn}$ autoregressively for $\mathbf{y}_\mathrm{syn}$.
\item Feed the reference tokens $\boldsymbol{\mu}_\mathrm{per}$, generated tokens $\boldsymbol{\mu}_\mathrm{syn}$, reference feature $\boldsymbol{x}_\mathrm{per}$, and speaker embedding $\boldsymbol{v}$ to the conditional flow matching (CFM) model and generate speech features $\boldsymbol{x}_\mathrm{syn}$ with ten iterations.
\item Using the vocoder to synthesize a waveform from $\boldsymbol{x}_\mathrm{syn}$.
\end{enumerate}
To enhance the similarity between synthetic and real utterances, classifier-free guidance \cite{ho2022classifier} is applied for both the LLM and CFM. 
Details are available in the repository\footnote{https://github.com/FunAudioLLM/CosyVoice}.

\subsection{FIP with Synthetic Personal Data for Personalized ASR}
In FIP, a neural network is considered a smooth function $f(\boldsymbol{x};\boldsymbol{w})$ that maps an input vector $\boldsymbol{x}$ to an output vector $\mathbf{y}$, where $\boldsymbol{w}$ denotes the trainable weights. FIP hypothesizes that the output $f(\boldsymbol{x};\boldsymbol{w})$ of a given neural network with a small weight perturbation $\mathrm{d}\boldsymbol{w}$ can be approximated as follows:
\begin{equation}
    f(\boldsymbol{x};\boldsymbol{w}+\mathrm{d}\boldsymbol{w})\approx f(\boldsymbol{x};\boldsymbol{w})+\mathbf{J}_{\boldsymbol{w}}\mathrm{d}\boldsymbol{w}
\end{equation}
where $\mathbf{J}_{\boldsymbol{w}}$ is the Jocabian of $f(\boldsymbol{x};\boldsymbol{w})$. Using this approximation, we can obtain the total difference between the outputs of two nearby networks with infinitesimal changes $\mathrm{d}\boldsymbol{w}$:
\begin{equation}
    \left| f(\boldsymbol{x};\boldsymbol{w}+\mathrm{d}\boldsymbol{w}) - f(\boldsymbol{x};\boldsymbol{w}) \right|^2 = \left| <\mathrm{d}\boldsymbol{w}, \mathrm{d}\boldsymbol{w}>_{\boldsymbol{r}_{w} } \right|^2
\end{equation}
where $\boldsymbol{r}_{w} = \mathbf{J}_{\boldsymbol{w}}(\boldsymbol{x})^\mathrm{T}\mathbf{J}_{\boldsymbol{w}}(\boldsymbol{x})$ is the metric tensor that features the weight space as a Riemannian manifold. FIP hypothesizes that there is a functionally invariant path $\boldsymbol{\psi}=\{\boldsymbol{w}_k\}$ in the Riemannian manifold that minimizes the loss function $L$ of a new task and output changes on the old task at the same time:
\begin{equation}   
    \mathrm{d}\boldsymbol{w}^*_k = \text{arg}\min_{\mathrm{d}\boldsymbol{w}_k}\left( \left< \mathrm{d}\boldsymbol{w}_k, \frac{\partial L}{\partial \boldsymbol{w}_{k}} \right>  + \beta \left< \mathrm{d}\boldsymbol{w}_k, \mathrm{d}\boldsymbol{w}_k \right>_{\boldsymbol{r}_{\boldsymbol{w}_k}} \right)
\label{eq:fip-org}
\end{equation}
where $\beta$ weighs the relative contribution of the two terms. 

\begin{figure}[t!]
  \centering
  \includegraphics[width=0.7\linewidth]{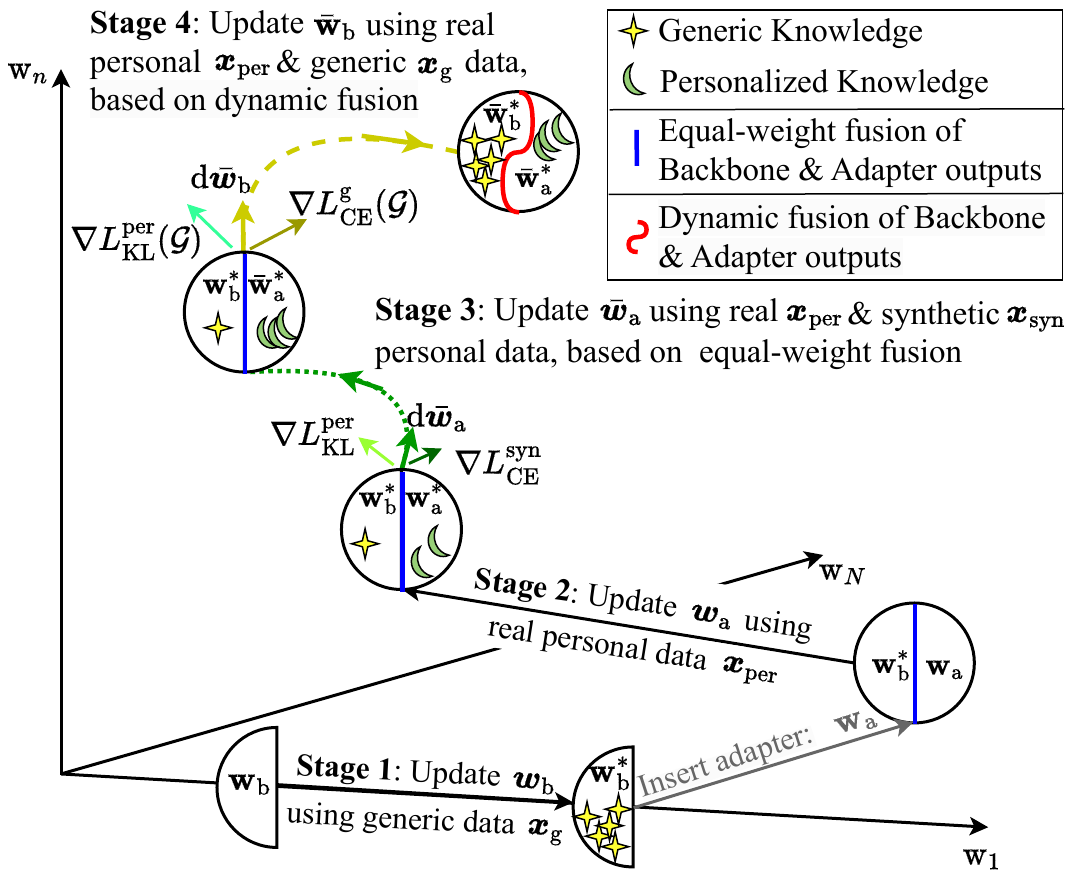}
  \caption{\scriptsize Schematic of KDFIP construction in the weight space $(\mathrm{w}_1$,$\mathrm{w}_{n}\dots\mathrm{w}_N)$ of ASR models for sequential training on personal and generic data, where the generic module $\mathbf{w}_\mathrm{b}$ and personalized module $\mathbf{w}_\mathrm{a}$ correspond to $(\mathrm{w}_1$,$\dots$,$\mathrm{w}_n)$ and $(\mathrm{w}_{n+1}$,$\dots$,$\mathrm{w}_N)$, respectively. The spherical and hemispherical shapes represent sets of model parameters at each stage. $\nabla L$ and $\mathrm{d}\boldsymbol{w}$ denote the gradient with respect to $L$ and the perturbation in the weights $\boldsymbol{w}$, respectively.}
  \label{fig:path}
\vspace{-0.5cm}
\end{figure}

To improve personalization using synthetic personal data while avoiding forgetting generic and real knowledge, it is reasonable to introduce FIP to adapt the generic ASR model for the target speaker.
Initially, given the generic ASR dataset $\{\boldsymbol{x}_\mathrm{g}, \mathbf{y}_\mathrm{g}\}$, we train the ASR model $f(\boldsymbol{x}_\mathrm{g};\boldsymbol{w}_\mathrm{b})$ to learn generic knowledge by minimizing the cross-entropy (CE) loss:
\begin{equation}
    \mathbf{w}_\mathrm{b}^* = \text{arg}\min_{\boldsymbol{w}_\mathrm{b}}\text{CE}\big(f(\boldsymbol{x}_\mathrm{g};\boldsymbol{w}_\mathrm{b}),~\mathbf{y}_\mathrm{g}\big)
\end{equation}
where the parameters of the generic ASR model are denoted as $\boldsymbol{w}_\mathrm{b}$, which serves as the backbone model in KDFIP. This training process corresponds to ``Stage~1'' in Fig.~\ref{fig:path}. As in Eq.~(\ref{eq:fip-org}), the optimization problem of FIP can be addressed by iteratively minimizing a hybrid loss function $\mathcal{L}$ that consists of the classification loss for personalized ASR tasks and the distance in the output space of networks for generic ASR tasks:
\begin{align}
    \mathcal{L} &=  \text{CE}\big(f\big(\{\boldsymbol{x}_\mathrm{per},\boldsymbol{x}_\mathrm{syn}\};\boldsymbol{w}_{\text{{\tiny FIP}}}\big),~\{\mathbf{y}_\mathrm{per},\mathbf{y}_\mathrm{syn}\}\big) \notag \\&+ \beta~\text{KL}\big(f(\boldsymbol{x}_\mathrm{g};\mathbf{w}_\mathrm{b}^*),~f(\boldsymbol{x}_\mathrm{g};\boldsymbol{w}_{\text{{\tiny FIP}}})\big) \label{eq:fip} \tag{5}
\end{align}
where the trainable parameters $\boldsymbol{w}_{\text{{\tiny FIP}}}$ are initialized from $\mathbf{w}_\mathrm{b}^*$, and the KL divergence measures the distance between the outputs of $f(\boldsymbol{x}_\mathrm{g};\mathbf{w}_\mathrm{b}^*)$ and $f(\boldsymbol{x}_\mathrm{g};\boldsymbol{w}_{\text{{\tiny FIP}}})$.
\subsection{Knowledge-Decoupled FIP}
\label{sec:KDFIP}
Although FIP enables adaptation to real and synthetic personal data without catastrophic forgetting, it still struggles to simultaneously learn and balance personalized and generic knowledge.
To tackle this issue, we propose knowledge-decoupled FIP (KDFIP), which adopts the parameter-isolation strategy and applies the FIP loss to independently learn personalized and generic knowledge.
Following PEFT, we introduce additional adapters to each layer in the backbone encoder, with the hidden states of the backbone and adapter equally weighted. Given the limited amount of real personal data, augmenting it with synthetic personal data is a viable approach. Thus, the adapter is trained by minimizing the CE loss on both synthetic and real personal data:
\begin{equation}
\footnotesize 
\begin{aligned}
  \mathbf{\hat{w}}_\mathrm{a}=\text{arg}\min_{\boldsymbol{w}_\mathrm{a}}\text{CE}\big(f\big(\{\boldsymbol{x}_\mathrm{per},\boldsymbol{x}_\mathrm{syn}\};\{\mathbf{w}_\mathrm{b}^*,\boldsymbol{w}_\mathrm{a}\}\big),~\{\mathbf{y}_\mathrm{per},\mathbf{y}_\mathrm{syn}\}\big)\label{eq:pga_adapter} 
\end{aligned} \tag{6}
\end{equation}
where $f(\cdot;\{\mathbf{w}_\mathrm{b}^*,\boldsymbol{w}_\mathrm{a}\})$ denotes the ASR model equipped with adapters parameterized by $\boldsymbol{w}_\mathrm{a}$. However, synthetic personal data may contain content or prosody errors due to language-model hallucinations, potentially impairing the training of adapters. The key challenge lies in learning from synthetic personal data without forgetting real personalized knowledge, which can be achieved by solving the \textbf{personalized functionality invariance} problem. In KDFIP, the adapter is first trained on personal data $\boldsymbol{x}_\mathrm{per}$ to acquire real personalized knowledge as illustrated in ``Stage 2'' of Fig.~\ref{fig:path}:
\begin{align}
    \mathbf{w}_\mathrm{a}^* = \text{arg}\min_{\boldsymbol{w}_\mathrm{a}}\text{CE}\big(f(\boldsymbol{x}_\mathrm{per};\{\mathbf{w}_\mathrm{b}^*,\boldsymbol{w}_\mathrm{a}\}), \mathbf{y}_\mathrm{per}\big)\label{eq:kdfip_adapter} \tag{7}
\end{align}
Then, the new adapter $\bar{\boldsymbol{w}}_\mathrm{a}$, initialized from the original $\mathbf{w}_\mathrm{a}^*$, is iteratively updated along personalization-invariant paths during ``Stage~3'' in Fig.~\ref{fig:path}, adapting to synthetic data without forgetting the real personalized knowledge contained in $\mathbf{w}_\mathrm{a}^*$:
\begin{align}
      \bar{\mathbf{w}}_\mathrm{a}^* &= \text{arg}\min_{\bar{\boldsymbol{w}}_\mathrm{a}}(L_\mathrm{CE}^\mathrm{syn} + L_\mathrm{KL}^\mathrm{per}) \notag \\
      &=\text{arg}\min_{\bar{\boldsymbol{w}}_\mathrm{a}}\big(\text{CE}(f(\boldsymbol{x}_\mathrm{syn};\{\mathbf{w}_\mathrm{b}^*,\bar{\boldsymbol{w}}_\mathrm{a}\}), \mathbf{y}_\mathrm{syn}) \notag \\&+ \beta~\text{KL}(f(\boldsymbol{x}_\mathrm{per};\{\mathbf{w}_\mathrm{b}^*,\mathbf{w}_\mathrm{a}^*\}), f(\boldsymbol{x}_\mathrm{per};\{\mathbf{w}_\mathrm{b}^*,\bar{\boldsymbol{w}}_\mathrm{a}\}))\big) \label{eq:stage3} \tag{8}
  \end{align}
 
The integration of personalized knowledge through the inserted adapters may influence the general recognition capability of the entire network. To restore generalization capabilities, it is necessary to dynamically fuse the learned knowledge. In PGA, a gating function $\mathcal{G}(\boldsymbol{x},\mathcal{X}_\mathrm{per})$ is utilized to control the proportion of adapter outputs $\boldsymbol{H}_{\mathrm{a}}$ added to backbone outputs $\boldsymbol{H}_{\mathrm{b}}$, based on the personality similarity between the input $\boldsymbol{x}$ and target speaker speeches $\mathcal{X}_\mathrm{per}=\{\boldsymbol{x}_\mathrm{per}\}$:
\begin{equation}
\begin{aligned}
    \boldsymbol{H} = \boldsymbol{H}_{\mathrm{b}} + \mathcal{G}(\boldsymbol{x},\mathcal{X}_\mathrm{per}) \cdot \boldsymbol{H}_{\mathrm{a}}
\end{aligned} \tag{9}
\end{equation}
According to PGA, the backbone is retrained to minimize the CE loss on generic $\boldsymbol{x}_{\mathrm{g}}$, personal $\boldsymbol{x}_{\mathrm{per}}$, and synthetic data $\boldsymbol{x}_{\mathrm{syn}}$:
  \begin{align}
\mathbf{w}_\mathrm{b}^{\text{{\tiny PGA}}}\!=\!\mathrm{arg}\min_{\boldsymbol{w}_{\mathrm{b}}}\big(\mathrm{CE}(&f(\{\boldsymbol{x}_\mathrm{g},\boldsymbol{x}_\mathrm{per},\boldsymbol{x}_\mathrm{syn}\};\{\boldsymbol{w}_\mathrm{b},\mathbf{\hat{w}}_\mathrm{a},\mathcal{G}\}), \notag \\ 
&\{\mathbf{y}_\mathrm{g},\mathbf{y}_\mathrm{per},\mathbf{y}_\mathrm{syn}\})\big) \label{eq:pga}  \tag{10}
  \end{align}
where $\boldsymbol{w}_{\mathrm{b}}$ is initialized from the backbone $\boldsymbol{w}_{\mathrm{b}}^*$. However, the spectral discrepancy between synthetic and real personal data hinders the gating function from assigning high gating scores to synthetic data, indicating that such data should be excluded from Eq.~(\ref{eq:pga}). Nevertheless, optimizing the backbone by minimizing CE loss on real personal data $\boldsymbol{x}_{\mathrm{per}}$ and generic data $\boldsymbol{x}_{\mathrm{g}}$ is not applicable after ``Stage~3'', as it may compromise the personalized knowledge acquired from synthetic data during ``Stage~3''.
We aim to recover the generalizability without forgetting the personalized knowledge acquired, which can be achieved by solving the problem of \textbf{personalized functionality invariance with the gating mechanism}. More precisely, KDFIP searches for an updating path of backbone parameters that preserves personalized functionality when the gating score is high and minimizes the CE loss on generic data when the gating score is low.
Accordingly, new backbone parameters $\bar{\boldsymbol{w}}_\mathrm{b}$, initialized from $\mathbf{w}_{\mathrm{b}}^*$, are fine-tuned along gated personalization-invariance paths during ``Stage~4'' in Fig.~\ref{fig:path}:

\noindent\makebox[0pt][l]{%
\begin{minipage}{0.49\textwidth}
  \begin{flalign*}
\!\bar{\mathbf{w}}_\mathrm{b}^*\!&=\text{arg}\min_{\bar{\boldsymbol{w}}_\mathrm{b}}(L_\mathrm{CE}^\mathrm{g} + L_\mathrm{KL}^\mathrm{per}) \notag\\
    &=\text{arg}\min_{\bar{\boldsymbol{w}}_\mathrm{b}}\big(\text{CE}(f(\boldsymbol{x}_\mathrm{g};\{\bar{\boldsymbol{w}}_\mathrm{b},\bar{\mathbf{w}}_\mathrm{a}^*,\mathcal{G}\}),~ \mathbf{y}_\mathrm{g}) \!\notag\\&+\!\beta~\text{KL}(f(\boldsymbol{x}_\mathrm{per};\{\mathbf{w}_\mathrm{b}^*,\bar{\mathbf{w}}_\mathrm{a}^*\}), f(\boldsymbol{x}_\mathrm{per};\{\bar{\boldsymbol{w}}_\mathrm{b},\bar{\mathbf{w}}_\mathrm{a}^*,\mathcal{G}\}))\big) \tag{11}
\end{flalign*} 
\end{minipage}
}
\vspace{0.2cm}

Upon convergence, the final model, $f(\cdot;\{\bar{\mathbf{w}}_\mathrm{b}^*, \bar{\mathbf{w}}_\mathrm{a}^*, \mathcal{G}\})$, restores the generalizability of models on non-target speaker input while improving the personalization for the target speaker. In summary, KDFIP fully exploits synthetic data to augment personalized knowledge and avoids catastrophic forgetting of generic knowledge by integrating the FIP and PGA.

\section{Experiments}
\subsection{Experimental Setup} 
To evaluate the proposed KDFIP, we utilize three open-source transcribed audio datasets: KeSpeech \cite{DBLP:conf/nips/Tang0XSLZWTXZYL21}, MagicData-Sichuan\footnote{https://magichub.com/datasets/sichuan-dialect-scripted-speech-corpus-daily-use-sentence/}, and MagicData-Zhengzhou\footnote{https://magichub.com/datasets/zhengzhou-dialect-scripted-speech-corpus-daily-use-sentence/}. The text segments for synthetic data are selected from an internal dataset containing 130,000 sentences.
The KeSpeech corpus contains 895 hours of utterances in Mandarin and its eight sub-dialects. According to the official data split, it is used as both the training dataset and the generic test set.
As the unseen speaker corpora, MagicData-Sichuan and MagicData-Zhengzhou, which belong to Southwestern Mandarin and Zhongyuan Mandarin, respectively, are employed as the adaptation data (ten minutes for each speaker) and the personal test set (twenty minutes for each speaker). Consistent with PGA \cite{gu24b_interspeech}, this letter uses data from six speakers, such as "s\_0001" and "z\_0011". The prefix "s" indicates Sichuan-accented Mandarin, while "z" denotes Zhengzhou-accented Mandarin.
For a fair comparison, we use the same text dataset to generate nearly 100 hours of speech for each target speaker.
We use the character error rate (CER) as the performance metric and also report the number of additional parameters per speaker and training steps to assess the model complexity and training cost of KDFIP.

\begin{table}[t!]
  \caption{\scriptsize Comparison of other speaker-adapted models: performance (CER, \%), parameter number, and training steps.}
  \label{tab:methodcomparition}
  \centering
  \scriptsize
  \setlength{\tabcolsep}{0.4mm}
  \renewcommand{\arraystretch}{1.0}
  
  \begin{tabular}{l|c|c|c|c|c|c}
    \hline 
    \multirow{2}{*}{Exps.}&\multirow{2}{*}{Model} & \multirow{1}{*}{Adaptation} & \multirow{1}{*}{Generic} & \multirow{1}{*}{Personal} & \multirow{1}{*}{\# Parameter} &  \multirow{2}{*}{Steps} \\
    && Data& test set&test set&per speaker&\\ 
    \hline
    Exp.~1&Base~(Stage~1) & \multirow{2}{*}{N/A} &$12.69$ & $19.06$ &$0$& $-$ \\
    Exp.~2&SAT & & $12.70$ & $18.00$ &$0$& $-$ \\
    \hline
    Exp.~3&FT &\multirow{3}{*}{$\mathrm{D_{per}}$}& $13.14$ & $17.82$ & $0$& $-$ \\
    Exp.~4& Adapter~(Stage~2)& & $20.27$ & $14.60$ & $1.9$\text{M} & $20$ \\
    Exp.~5&PGA && $12.96$  & $14.94$ & $1.9$\text{M} & $-$\\
    \hline
    Exp.~6&FT &\multirow{3}{*}{$\mathrm{D_{per}+D_{syn}}$}& $20.81$ & $13.79$ & $0$& $39.9$\text{K} \\
    Exp.~7 & Adapter& & $27.47$ & $14.57$ & $1.9$\text{M} &$8.0$\text{K} \\ 
    Exp.~8&PGA && $13.41$  & $14.38$ &$1.9$\text{M} & $13.1$\text{K} \\
    \hline
    Exp.~9&FIP &\multirow{3}{*}{$\mathrm{D_{per}+D_{syn}}$}& $13.93$  & $14.17$ & $0$& $39.9$\text{K} \\
    Exp.~10 & Adapter-FIP~(Stage~3) & & $20.28$ & $\textbf{13.30}$ & $1.9$\text{M} & $2.4$\text{K} \\
    Exp.~11&KDFIP~(Stage~4)&  & $13.16$ & $\text{13.46}$ & $1.9$\text{M} &$15.5$\text{K}\\
    \hline
  \end{tabular}
  \vspace{-0.6cm}
\end{table}

The base (backbone) model, the gating function model, the adapter module, and the TTS model are pre-trained and publicly released \cite{gu23_interspeech, gu24b_interspeech,du2024cosyvoice2}, thus we use them directly in our experiments.
The hyperparameter $\beta$ is set to 0.01 in all experiments. The FIP ASR model is trained for 50 epochs. For the KDFIP setting, the adapter module is trained for three epochs with a learning rate of 0.0015, and the backbone is trained for three epochs with a learning rate of 5e-7.

\subsection{Experimental results}
Table~\ref{tab:methodcomparition} presents the effectiveness of synthetic data augmentation and compares the proposed KDFIP with other speaker adaptation methods in terms of CER on generic and personal test sets. $\mathrm{D_{per}}$ means ten minutes real personal data and $\mathrm{D_{syn}}$ refers to 100 hours synthetic personal data. Notably, almost all models, except for SAT, share the same pre-trained backbone model. In speaker adaptive training (SAT), the speaker embedding is concatenated to the input of the conformer block to obtain speaker-dependent ASR models.
Although this method does not require retraining, SAT ASR models depend on an external speaker embedding extractor and achieve only limited performance improvement on the personal test set. 

As shown in the Table \ref{tab:methodcomparition}, incorporating personal adaptation data into the ASR model training leads to better personalization performance.
It is a simple and effective strategy that directly fine-tunes (FT) the pre-trained generic ASR model with adaptation data \cite{DBLP:conf/interspeech/0028Y0021,yang2023text}.
Comparing the experiments Exp.~3 and Exp.~6, FT benefits more from the combination of synthetic and real personal data than from using only ten minutes of real personal data, demonstrating the effectiveness of synthetic personal data augmentation. In Exp.~7, the adapter is directly trained with CE loss on both synthetic and real personal data (Eq.~(\ref{eq:pga_adapter})). Based on this, the PGA model in Exp.~8 is trained according to Eq.~(\ref{eq:pga}). Notably, the difference between Exp.~4 and Exp.~7, and between Exp.~5 and Exp.~8 lies in the use of synthetic data. In addition, the FIP model is trained following Eq.~(\ref{eq:fip}), while the Adapter-FIP model is trained during ``Stage~3''. Comparing Exp.~4 and Exp.~7, we observe that the adapter fails to benefit from the synthetic data, as Exp.~7 shows degraded performance on the generic dataset. This is possibly due to the adapter being sensitive to phonetic errors introduced by hallucinations of the TTS model. As a result, although the PGA model (Exp.~8) maintains generalizability when synthetic data is used, it performs worse than FT (Exp.~6) and KDFIP on personal test sets. 

The FIP model struggles to learn from the generic data, real and synthetic personal data simultaneously. Without knowledge decoupling, it becomes difficult for FIP to balance multiple types of knowledge, leading to suboptimal performance on both generic and personal test sets. Compared with the base model, KDFIP achieves the highest relative improvement of 29.38\% among the aforementioned methods on personal test sets, with only a slight deterioration in generalizability. In addition, KDFIP requires only a limited number of training steps for adaptation.

\begin{table}[t!]
  \caption{\scriptsize Ablation study on ``Stage 3'' of the proposed KDFIP using synthetic data from target (s\_1100) and non-target speakers (CER, \%).}
  \label{tab:spkcross}
  \scriptsize
  \centering
  \setlength{\tabcolsep}{2.0mm}
  \renewcommand{\arraystretch}{1.5}
  \begin{tabular}{c|c|c|c}
    \hline 
    Adaptation Data & CER & Adaptation Data & CER \\     
    \hline
    $\mathrm{D_{per}^{s\_1100}}+\text{None~~~~}$ & $19.67$ & $\mathrm{D_{per}^{s\_1100}+D_{syn}^{s\_1100}}$ & $\textbf{17.89}$ \\
    $\mathrm{D_{per}^{s\_1100}+D_{syn}^{s\_0001}}$ & $18.48$ & $\mathrm{D_{per}^{s\_1100}+D_{syn}^{s\_1106}}$ & $19.03$\\
    $\mathrm{D_{per}^{s\_1100}+D_{syn}^{z\_0011}}$ & $19.85$ & $\mathrm{D_{per}^{s\_1100}+D_{syn}^{z\_1196}}$ & $19.30$\\
    \hline
  \end{tabular}
  \vspace{-0.5cm}
\end{table}
\subsection{Ablation study}
To examine the effect of the acoustic information from target speakers, we conduct an ablation experiment by combining ten minutes of real personal data from speaker ``s\_1100'' with 100 hours of synthetic personal data from other speakers in ``Stage~3'' of KDFIP, based on the ``Stage~2'' model of speaker ``s\_1100''. As shown in Table~\ref{tab:spkcross}, although the synthetic data from other speakers shares the same textual context as the target speaker, the ASR model performs better on the target speaker when the corresponding synthetic data is involved. We believe this improvement is related to the powerful capabilities of large TTS models.

\begin{figure}[t!]
  \scriptsize
  \centering
  \includegraphics[width=0.8\linewidth]{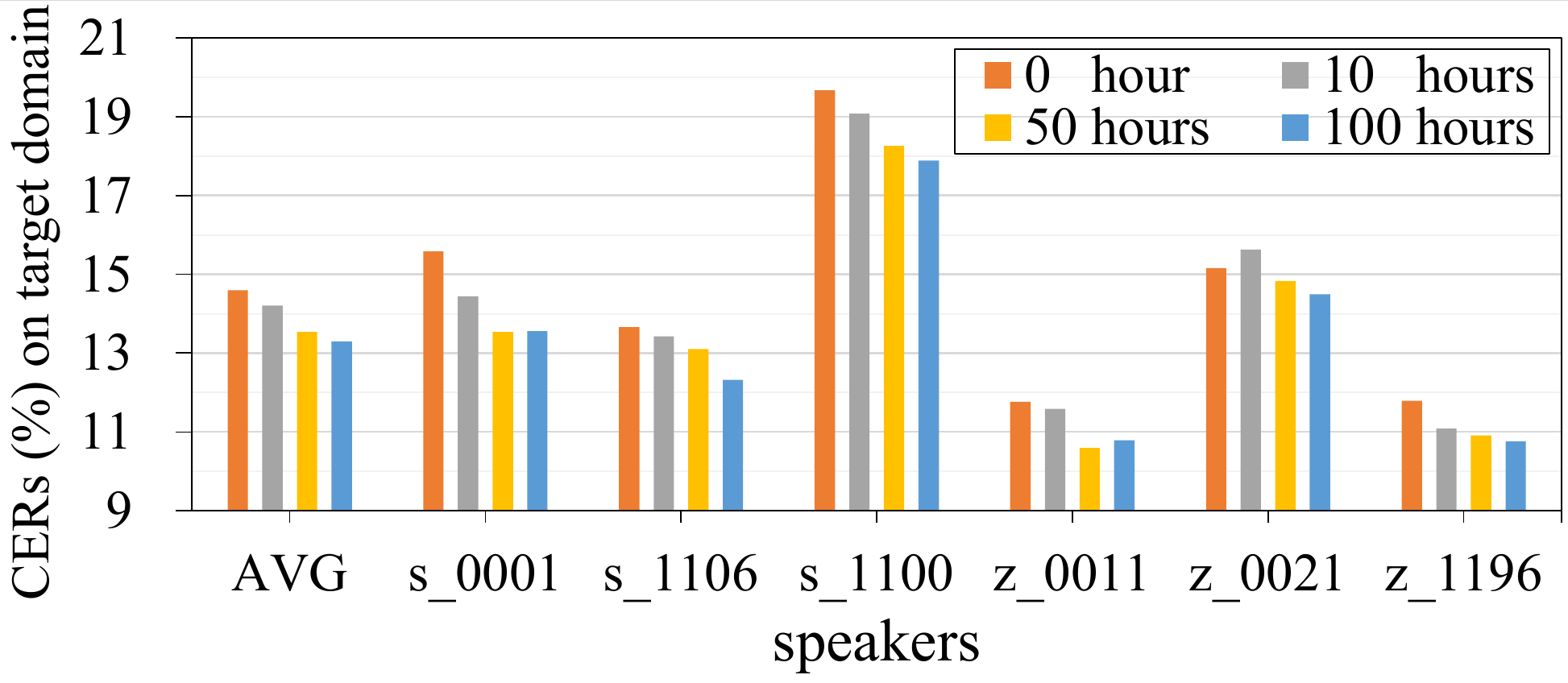}
  \caption{\scriptsize Ablation on the duration of synthetic personal data in ``Stage~3'' of KDFIP.}
  \label{fig:duration}
  \vspace{-0.3cm}
\end{figure}

Fig.~\ref{fig:duration} illustrates the influence of the synthetic personal data duration in ``Stage~3''. Overall, increasing the amount of synthetic personal data leads to performance improvements. However, there is no clear difference between 50 and 100 hours of adaptation data for some speakers, such as ``s\_0001'' and ``z\_1196''. This may be due to the underrepresentation of the target speaker's voice characteristics in the limited ten-minute real personal dataset, causing the TTS model to generate data with insufficient speaker profiles.
\subsection{Hyperparameter Tuning}

\begin{figure}[t!]
  \centering
  \includegraphics[width=0.8\linewidth]{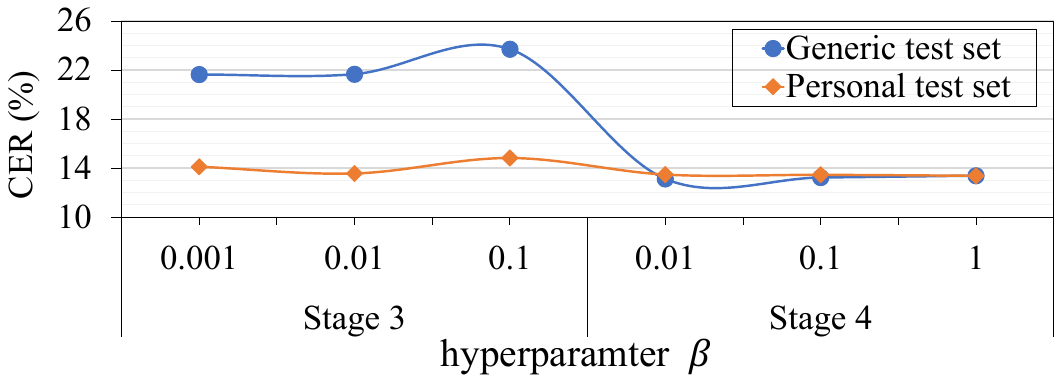}
  \caption{Interpolated hyperparameter tuning of KDFIP.}
  \label{fig:hyper}
  \vspace{-0.5cm}
\end{figure}

The interpolated hyperparameter $\beta$, denoted in the method section, is tuned to assist model training. Fig.~\ref{fig:hyper} shows the CERs on both generic and personal test sets of ``Stage 3'' and ``Stage 4'' as $\beta$ varies. The figure indicates that the ASR model is robust to the change of $\beta$ during ``Stage 4'', while it is sensible to larger $\beta$, such as 0.1, during ``Stage 3''. In this letter, $\beta$ is set to 0.01 for all experiments.

\section{Conclusion}
It is difficult for personalized ASR models to learn and balance the synthetic knowledge, real personalized knowledge, and generic knowledge when amounts of synthetic personal data are involved in the model training.
In this letter, we propose KDFIP to decouple the learning of three types of knowledge into personalization and generalizability adaptations, significantly improving personalized ASR without compromising generalizability. Experimental results on three open-source datasets demonstrate the effectiveness of KDFIP and synthetic personal data augmentation. Additionally, the ablation study indicates that the target speaker's voice characteristics are a critical factor in data augmentation using large-scale TTS models.

\bibliographystyle{IEEEtran}
\bibliography{mybib}

\begin{thebibliography}{10}
\providecommand{\url}[1]{#1}
\csname url@samestyle\endcsname
\providecommand{\newblock}{\relax}
\providecommand{\bibinfo}[2]{#2}
\providecommand{\BIBentrySTDinterwordspacing}{\spaceskip=0pt\relax}
\providecommand{\BIBentryALTinterwordstretchfactor}{4}
\providecommand{\BIBentryALTinterwordspacing}{\spaceskip=\fontdimen2\font plus
\BIBentryALTinterwordstretchfactor\fontdimen3\font minus
  \fontdimen4\font\relax}
\providecommand{\BIBforeignlanguage}[2]{{%
\expandafter\ifx\csname l@#1\endcsname\relax
\typeout{** WARNING: IEEEtran.bst: No hyphenation pattern has been}%
\typeout{** loaded for the language `#1'. Using the pattern for}%
\typeout{** the default language instead.}%
\else
\language=\csname l@#1\endcsname
\fi
#2}}
\providecommand{\BIBdecl}{\relax}
\BIBdecl

\bibitem{DBLP:journals/spl/LeeMKSC25}
M.~Lee, J.~Mo, J.~Kang, J.~Son, and J.~Chang, ``Bayesian language model
  adaptation for personalized speech recognition,'' \emph{{SPL}}, vol.~32, pp.
  1620--1624, 2025.

\bibitem{DBLP:journals/taslp/DehakKDDO11}
N.~Dehak, P.~Kenny, R.~Dehak, P.~Dumouchel, and P.~Ouellet, ``Front-end factor
  analysis for speaker verification,'' \emph{TASLP}, vol.~19, pp. 788--798,
  2011.

\bibitem{DBLP:conf/icassp/SnyderGSPK18}
D.~Snyder, D.~Garcia{-}Romero, G.~Sell, D.~Povey, and S.~Khudanpur,
  ``X-vectors: Robust {DNN} embeddings for speaker recognition,'' in
  \emph{{ICASSP}}, 2018, pp. 5329--5333.

\bibitem{xue2014fast}
S.~Xue, O.~Abdel-Hamid, H.~Jiang, L.~Dai, and Q.~Liu, ``Fast adaptation of deep
  neural network based on discriminant codes for speech recognition,''
  \emph{TASLP}, vol.~22, pp. 1713--1725, 2014.

\bibitem{DBLP:conf/interspeech/DelcroixWOKN18}
M.~Delcroix, S.~Watanabe, A.~Ogawa, S.~Karita, and T.~Nakatani, ``Auxiliary
  feature based adaptation of end-to-end {ASR} systems,'' in
  \emph{{INTERSPEECH}}, 2018, pp. 2444--2448.

\bibitem{li24ka_interspeech}
S.~Li, D.~Wei, H.~Shang, J.~Guo, Z.~Li \emph{et~al.}, ``Speaker-smoothed knn
  speaker adaptation for end-to-end asr,'' in \emph{{INTERSPEECH}}, 2024, pp.
  2390--2394.

\bibitem{saon2013speaker}
G.~Saon, H.~Soltau, D.~Nahamoo, and M.~Picheny, ``Speaker adaptation of neural
  network acoustic models using i-vectors,'' in \emph{ASRU}, 2013, pp. 55--59.

\bibitem{DBLP:conf/icassp/SeniorL14}
A.~W. Senior and I.~L{\'{o}}pez{-}Moreno, ``Improving {DNN} speaker
  independence with i-vector inputs,'' in \emph{{ICASSP}}, 2014, pp. 225--229.

\bibitem{DBLP:conf/interspeech/ZeineldeenXLSN22}
M.~Zeineldeen, J.~Xu, C.~L{\"{u}}scher, R.~Schl{\"{u}}ter, and H.~Ney,
  ``Improving the training recipe for a robust conformer-based hybrid model,''
  in \emph{{INTERSPEECH}}, 2022, pp. 1036--1040.

\bibitem{sari2020unsupervised}
L.~Sar{\i}, N.~Moritz, T.~Hori, and J.~Le~Roux, ``Unsupervised speaker
  adaptation using attention-based speaker memory for end-to-end asr,'' in
  \emph{ICASSP}, 2020, pp. 7384--7388.

\bibitem{fan2019speaker}
Z.~Fan, J.~Li, S.~Zhou, and B.~Xu, ``Speaker-aware speech-transformer,'' in
  \emph{ASRU}, 2019, pp. 222--229.

\bibitem{zhao2020speech}
Y.~Zhao, C.~Ni, C.-C. Leung, S.~R. Joty, E.~S. Chng, and B.~Ma, ``Speech
  transformer with speaker aware persistent memory,'' in \emph{INTERSPEECH},
  2020, pp. 1261--1265.

\bibitem{wan2020speaker}
G.~Wan, J.~Pan, Q.~Wang, J.~Gao, and Z.~Ye, ``Speaker adaptive training for
  speech recognition based on attention-over-attention mechanism,'' in
  \emph{INTERSPEECH}, 2020, pp. 1251--1255.

\bibitem{DBLP:conf/interspeech/0028Y0021}
Y.~Huang, G.~Ye, J.~Li, and Y.~Gong, ``Rapid speaker adaptation for conformer
  transducer: Attention and bias are all you need,'' in \emph{INTERSPEECH},
  2021, pp. 1309--1313.

\bibitem{gu23_interspeech}
Y.~Gu, Z.~Du, S.~Zhang, Q.~Chen, and J.~Han, ``{Personality-aware Training
  based Speaker Adaptation for End-to-end Speech Recognition},'' in
  \emph{INTERSPEECH}, 2023, pp. 1249--1253.

\bibitem{swietojanski2016learning}
P.~Swietojanski, J.~Li, and S.~Renals, ``Learning hidden unit contributions for
  unsupervised acoustic model adaptation,'' \emph{TASLP}, vol.~24, pp.
  1450--1463, 2016.

\bibitem{wang2017unsupervised}
Z.-Q. Wang and D.~Wang, ``Unsupervised speaker adaptation of batch normalized
  acoustic models for robust asr,'' in \emph{ICASSP}, 2017, pp. 4890--4894.

\bibitem{xie2021bayesian}
X.~Xie, X.~Liu, T.~Lee, and L.~Wang, ``Bayesian learning for deep neural
  network adaptation,'' \emph{TASLP}, vol.~29, pp. 2096--2110, 2021.

\bibitem{deng23b_interspeech}
J.~Deng, G.~Li, X.~Xie, Z.~Jin, M.~Cui, T.~Wang, S.~Hu, M.~Geng, and X.~Liu,
  ``{Factorised Speaker-environment Adaptive Training of Conformer Speech
  Recognition Systems},'' in \emph{INTERSPEECH}, 2023, pp. 3342--3346.

\bibitem{gu24b_interspeech}
Y.~Gu, Z.~Du, S.~Zhang, jiqing Han, and Y.~He, ``Personality-memory gated
  adaptation: An efficient speaker adaptation for personalized end-to-end
  automatic speech recognition,'' in \emph{{INTERSPEECH}}, 2024, pp.
  2870--2874.

\bibitem{bell2020adaptation}
P.~Bell, J.~Fainberg, O.~Klejch, J.~Li, S.~Renals, and P.~Swietojanski,
  ``Adaptation algorithms for neural network-based speech recognition: An
  overview,'' \emph{IEEE Open Journal of Signal Processing}, vol.~2, pp.
  33--66, 2020.

\bibitem{houlsby2019parameter}
N.~Houlsby, A.~Giurgiu, S.~Jastrzebski, B.~Morrone, Q.~De~Laroussilhe
  \emph{et~al.}, ``Parameter-efficient transfer learning for nlp,'' in
  \emph{ICML}, 2019, pp. 2790--2799.

\bibitem{DBLP:conf/iclr/HuSWALWWC22}
E.~J. Hu, Y.~Shen, P.~Wallis, Z.~Allen{-}Zhu, Y.~Li, S.~Wang, L.~Wang, and
  W.~Chen, ``Lora: Low-rank adaptation of large language models,'' in
  \emph{{ICLR}}, 2022.

\bibitem{hu2023llm}
Z.~Hu, Y.~Lan, L.~Wang, W.~Xu, E.-P. Lim, R.~Ka-Wei~Lee, L.~Bing, and S.~Poria,
  ``Llm-adapters: An adapter family for parameter-efficient fine-tuning of
  large language models,'' in \emph{{EMNLP}}, 2023, pp. 5254–--5276.

\bibitem{li2023evaluating}
Y.~Li, A.~Mehrish, R.~Bhardwaj, N.~Majumder, B.~Cheng \emph{et~al.},
  ``Evaluating parameter-efficient transfer learning approaches on sure
  benchmark for speech understanding,'' in \emph{ICASSP}, 2023, pp. 1--5.

\bibitem{DBLP:conf/acl/WangLJWWJCHWSZ23}
Z.~Wang, Y.~Liu, T.~Ji, X.~Wang, Y.~Wu, C.~Jiang, Y.~Chao, Z.~Han, L.~Wang,
  X.~Shao, and W.~Zeng, ``Rehearsal-free continual language learning via
  efficient parameter isolation,'' in \emph{{ACL} {(1)}}.\hskip 1em plus 0.5em
  minus 0.4em\relax Association for Computational Linguistics, 2023, pp.
  10\,933--10\,946.

\bibitem{huang2020using}
Y.~Huang, L.~He, W.~Wei, W.~Gale, J.~Li, and Y.~Gong, ``Using personalized
  speech synthesis and neural language generator for rapid speaker
  adaptation,'' in \emph{ICASSP}, 2020, pp. 7399--7403.

\bibitem{yang2023text}
K.~Yang, T.-Y. Hu, J.-H.~R. Chang, H.~S. Koppula, and O.~Tuzel, ``Text is all
  you need: Personalizing asr models using controllable speech synthesis,'' in
  \emph{ICASSP}, 2023, pp. 1--5.

\bibitem{DBLP:conf/icassp/KimLC24}
D.~Kim, J.~Lee, and J.~Chang, ``Text-only unsupervised domain adaptation for
  neural transducer-based {ASR} personalization using synthesized data,'' in
  \emph{{ICASSP}}, 2024, pp. 11\,131--11\,135.

\bibitem{du2024cosyvoice}
Z.~Du, Q.~Chen, S.~Zhang, K.~Hu, H.~Lu \emph{et~al.}, ``Cosyvoice: A scalable
  multilingual zero-shot text-to-speech synthesizer based on supervised
  semantic tokens,'' \emph{arXiv preprint arXiv:2407.05407}, 2024.

\bibitem{du2024cosyvoice2}
Z.~Du, Y.~Wang, Q.~Chen, X.~Shi, X.~Lv \emph{et~al.}, ``Cosyvoice 2: Scalable
  streaming speech synthesis with large language models,'' \emph{arXiv preprint
  arXiv:2412.10117}, 2024.

\bibitem{du2025cosyvoice}
Z.~Du, C.~Gao, Y.~Wang, F.~Yu, T.~Zhao \emph{et~al.}, ``Cosyvoice 3: Towards
  in-the-wild speech generation via scaling-up and post-training,'' \emph{arXiv
  preprint arXiv:2505.17589}, 2025.

\bibitem{anastassiou2024seed}
P.~Anastassiou, J.~Chen, J.~Chen, Y.~Chen, Z.~Chen \emph{et~al.}, ``Seed-tts: A
  family of high-quality versatile speech generation models,'' \emph{arXiv
  preprint arXiv:2406.02430}, 2024.

\bibitem{le2023voicebox}
M.~Le, A.~Vyas, B.~Shi, B.~Karrer, L.~Sari \emph{et~al.}, ``Voicebox:
  Text-guided multilingual universal speech generation at scale,''
  \emph{Advances in neural information processing systems}, vol.~36, pp.
  14\,005--14\,034, 2023.

\bibitem{yang2025enhancing}
G.~Yang, F.~Yu, Z.~Ma, Z.~Du, Z.~Gao, S.~Zhang, and X.~Chen, ``Enhancing
  low-resource asr through versatile tts: Bridging the data gap,'' in
  \emph{{ICASSP}}, 2025, pp. 1--5.

\bibitem{liu2025mitigating}
\BIBentryALTinterwordspacing
C.~Liu, M.~Fang, P.~Zhang, W.~Zhou, J.~Gao, and J.~Han, ``Mitigating
  hallucinations in lm-based tts models via distribution alignment using
  gflownets,'' in \emph{EMNLP}, 2025, accepted for publication. [Online].
  Available: \url{https://arxiv.org/abs/2508.15442}
\BIBentrySTDinterwordspacing

\bibitem{vander2023using}
S.~Vander~Eeckt and H.~Van~Hamme, ``Using adapters to overcome catastrophic
  forgetting in end-to-end automatic speech recognition,'' in \emph{ICASSP},
  2023, pp. 1--5.

\bibitem{raghavan2024engineering}
G.~Raghavan, B.~Tharwat, S.~N. Hari, D.~Satani, R.~Liu, and M.~Thomson,
  ``Engineering flexible machine learning systems by traversing functionally
  invariant paths,'' \emph{Nature Machine Intelligence}, vol.~6, pp.
  1179--1196, 2024.

\bibitem{ho2022classifier}
J.~Ho and T.~Salimans, ``Classifier-free diffusion guidance,'' \emph{arXiv
  preprint arXiv:2207.12598}, 2022.

\bibitem{DBLP:conf/nips/Tang0XSLZWTXZYL21}
Z.~Tang, D.~Wang, Y.~Xu, J.~Sun, and et.al., ``Kespeech: An open source speech
  dataset of mandarin and its eight subdialects,'' in \emph{NeurIPS Datasets
  and Benchmarks}, 2021.

\end{thebibliography}

\end{document}